\documentclass[useAMS,usenatbib]{mn2e}

\usepackage{amssymb, amsmath}
\usepackage{epsf}
\usepackage{bm}

\def\la{\; \raise0.3ex\hbox{$<$\kern-0.75em\raise-1.1ex\hbox{$\sim$}}\;}
\def\ga{\;  \raise0.3ex\hbox{$>$\kern-0.75em\raise-1.1ex\hbox{$\sim$}}\;}

\title[Velocity-dependent energy gaps]
{Velocity-dependent energy gaps and
dynamics of superfluid neutron stars}
\author[M. E. Gusakov, E. M. Kantor]
{M. E. Gusakov$^1$\thanks{gusakov@astro.ioffe.ru},
E.~M.~Kantor$^{1,2}$\thanks{kantor@mail.ioffe.ru}\\
$^1$Ioffe Physical-Technical Institute of the Russian Academy of
Sciences,
Polytekhnicheskaya 26, 194021 St.-Petersburg, Russia
\\
$^2$St.-Petersburg State Polytechnical University,
Polytekhnicheskaya 29, 195251 St.-Petersburg, Russia
}

\begin{document}

\date{Accepted 2012 xxxx. Received 2012 xxxx;
in original form 2012 xxxx}

\pagerange{\pageref{firstpage}--\pageref{lastpage}} \pubyear{2012}

\maketitle

\label{firstpage}

%
\begin{abstract}
We show that suppression of the baryon energy gaps,
caused by the relative motion of superfluid and normal liquid components,
can substantially influence dynamical properties and evolution of neutron stars.
This effect has been previously ignored in the neutron-star literature. 
\end{abstract}
%


\maketitle

\section{Introduction}

According to numerous microscopic calculations 
(e.g., \citealt*{yls99, ls01} and references therein), 
nucleons and hyperons
in the internal layers of neutron stars (NSs) can become superfluid
at temperatures $T \la 10^8 \div10^{10}$~K. 
Superfluidity has a strong impact on the thermal evolution
of NSs, their oscillations,  and (most probably) leads to 
such observational phenomena 
as glitches (\citealt*{ai75})
and pulsar spin precession (\citealt*{shaham77, lc02}). 
Recent real-time observations (\citealt*{hh10}) 
of a cooling NS in Cassiopea A supernova remnant 
give strong arguments 
that the star has superfluid core (\citealt{syhhp11,ppls11}).

The aim of this short note is to point out 
the importance 
of one effect related to superfluidity of baryons in NSs
that has usually been ignored in the NS literature.
In Sec.\ II we outline the effect.
In Sec.\ III we demonstrate its efficiency.
In Sec.\ IV we discuss possible 
consequences for the physics of NSs
and in Sec.\ V we conclude.
We use the system of units 
in which $k_{\rm B}=\hbar=1$.

\section{A simple problem and the proposed effect}

Let us consider a degenerate Fermi-liquid 
composed of identical particles of mass $m$.
Assume that they interact through a weakly attractive potential so that 
BCS theory (see, e.g., \citealt*{lp80}) is applicable.
Assume also that they pair 
(become superfluid) 
in the spin-singlet $^1S_0$ state
at temperatures $T$ below some critical temperature $T_{\rm c}$.
The role of elementary excitations in such superfluid Fermi-liquid
is played by Bogoliubov excitations (see, e.g., \citealt*{feynman72}).
In what follows, all equations will be
written in a reference frame in which 
the mean (hydrodynamic) velocity ${\pmb V}_{\rm q}$ 
of Bogoliubov excitations vanishes, ${\pmb V}_{\rm q}=0$
(i.e., normal liquid component is at rest). 

In the absence of superfluid current 
(when the superfluid velocity 
${\pmb V}_{\rm s}=0$)
the energy $E_{\pmb p}$ of a Bogoliubov excitation 
with momentum ${\pmb p}$
near the Fermi surface can be written as
\begin{equation}
E_{\pmb p}=\sqrt{v_{\rm F}^2 (|{\pmb p}|-p_{\rm F})^2 + \Delta^2},
\label{disp1}
\end{equation}
where $v_{\rm F}$ and $p_{\rm F}$ 
are the Fermi-velocity and Fermi-momentum, respectively;
and $\Delta$ is the energy gap, 
given by the standard equation (\citealt{lp80}),
\begin{equation}
1 = - V_0 \sum_{{\pmb p}} \frac{1- 2 f_{{\pmb p}}}{2 E_{{\pmb p}}},
\label{gap1}
\end{equation}
where $V_0$ is the (constant) pairing potential and
\begin{equation}
f_{\pmb p}=\frac{1}{{\rm e}^{E_{\pmb p}/T}+1}
\label{fp}
\end{equation}
is the Fermi-Dirac distribution function for Bogoliubov excitations.

If, however, the superfluid current 
is present (${\pmb V}_{\rm s}\neq 0$) 
then fermions pair with momenta $(-{\pmb p}+{\pmb Q}, {\pmb p}+{\pmb Q})$
rather than with $(-{\pmb p}, {\pmb p})$, 
and the total momentum of a Cooper pair is 
\begin{equation}
2 \,  {\pmb Q}= 2 \, m {\pmb V}_{\rm s}.
\label{Q}
\end{equation}

What will be the equation for the gap? 
The answer can be found in \cite{bardeen62} 
and is well known in the physics of superconductors.
Now, instead of Eq.\ (\ref{gap1}), one should write
\begin{equation}
1 = - V_0 \sum_{\pmb p} 
\frac{1-\mathcal{F}_{{\pmb p}+{\pmb Q}}-\mathcal{F}_{-{\pmb p}+{\pmb Q}}}{2 E_{\pmb p}}.
\label{gap2}
\end{equation}
Here $\mathcal{F}_{{\pmb p}+{\pmb Q}}$ 
is the distribution function for Bogoliubov 
excitations with momentum $({\pmb p}+{\pmb Q})$ 
in the system with non-zero ${\pmb V}_{\rm s}$,
\begin{equation}
\mathcal{F}_{{\pmb p}+{\pmb Q}}= \frac{1}{{\rm e}^{\mathfrak{E}_{{\pmb p}+{\pmb Q}}/T}+1}, 
\label{Fp2}
\end{equation}
where 
\begin{equation}
\mathfrak{E}_{{\pmb p}+{\pmb Q}} \approx \frac{{\pmb p} {\pmb Q}}{m} + E_{\pmb p}
\label{disp2}
\end{equation}
is the energy of a Bogoliubov excitation with momentum $({\pmb p}+{\pmb Q})$.
In Eq.\ (\ref{disp2}) we assumed $Q \ll p_{\rm F}$ 
which is true in all interesting cases 
(see, e.g., \citealt*{gh05,gkh09a,gkh09b}).

Eq.\ (\ref{gap2}) can be written in terms of the quantity $\Delta_0$, 
which is the energy gap at $T=0$ and ${\pmb Q}=0$.
It satisfies Eq.\ (\ref{gap1}) with $f_{\pmb p}=0$.
Using it, one can present Eq.\ (\ref{gap2}) in the form
\begin{equation}
\frac{p_{\rm F} m}{\pi^2} \,\, {\rm ln}\left(\frac{\Delta_0}{\Delta}\right) = 
\sum_{\pmb p} \frac{\mathcal{F}_{{\pmb p}+{\pmb Q}}+\mathcal{F}_{-{\pmb p}+{\pmb Q}}}{E_{\pmb p}}.
\label{gap3}
\end{equation}
The solution to this equation gives the gap $\Delta$ 
as a function of $T$ and $Q=|{\pmb Q}|$.

First consider two limiting cases 
in which $\Delta(T, \, Q)$ vanishes.

($i$) if ${Q}=0$ then $\Delta=0$ at 
\begin{equation}
T=T_{\rm c}\approx 0.567 \Delta_0 \,\,\,{\rm ( the \, well \, known \, BCS \, result);}
\label{Tc}
\end{equation}

($ii$) if $T=0$ then $\Delta=0$ at 
\begin{equation}
Q \equiv Q_{\rm cr \, 0}= \frac{\rm e}{2} \, \,\frac{\Delta_0 \, m}{p_{\rm F}}.
\label{Qcr}
\end{equation}
The latter result is less known but can be found, e.g., in \cite{alexandrov03}.
Notice that, the well-known Landau criterion 
for superfluidity breaking
gives $Q_{\rm cr \, 0}^{({\rm Landau})} =\Delta_0\,  m/p_{\rm F}$
and is not accurate for
a superfluid Fermi-liquid.
%
\begin{figure}
    \begin{center}
        \leavevmode
        \epsfxsize=2.0in \epsfbox{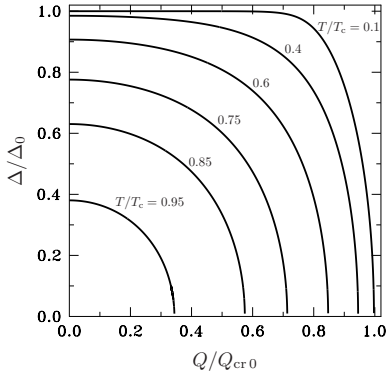}
    \end{center}
    \caption{The energy gap $\Delta$ (in units of $\Delta_0$)
    versus $Q=m V_{\rm s}$ [in units of $Q_{\rm cr \, 0}$, see Eq.\ (\ref{Qcr})] 
    for a set of temperatures
    $T/T_{\rm c}=0.1$, $0.4$, $0.6$, $0.75$, $0.85$, and $0.95$.
    }
    \label{Fig:gap_V}
\end{figure}
%

Some numerical solutions to Eq.\ (\ref{gap3}) 
are presented in Figs.\ 1 and 2.
Fig.\ 1 shows the gap $\Delta(T, \, Q)$ [in units of $\Delta_0$] 
versus momentum $Q$ [in units of $Q_{\rm cr \, 0}$]
for a set of temperatures $T/T_{\rm c}=0.1$, $0.4$, $0.6$, 
$0.75$, $0.85$, and $0.95$.
One sees that $\Delta$ is quite sensitive to variation 
of $Q = m \, V_{\rm s}$ 
as long as $T \ga 0.1 T_{\rm c}$.
Another important conclusion that can be drawn 
from Fig.\ 1 
is that (for a given $T$) 
the maximum critical momentum $Q_{\rm cr}$
strongly depends on temperature.

Fig.\ 2 illustrates this point more clearly.
In the left panel we plot $Q_{\rm cr}$ (in units of $Q_{\rm cr \, 0}$)
versus $T$ (in units of $T_{\rm c}$).
The right panel shows the same dependence $Q_{\rm cr}(T)$ but 
with $Q_{\rm cr}$ measured in units of 
\begin{equation}
Q_{\rm cr}^{({\rm app})}(T) \equiv \frac{{\rm e}}{2} \,\, \frac{\Delta(T,\, 0)\,  m}{p_{\rm F}}.
\label{Qcrapp}
\end{equation}
We see that 
$Q_{\rm cr}$ changes with $T$ in such a way that
$Q_{\rm cr}(T)/Q_{\rm cr}^{({\rm app})}(T)$ is roughly constant.
 
Therefore, 
the energy gap $\Delta$ can be a strong function of the momentum
${\pmb Q}=\, m\, {\pmb V}_{\rm s}$ or, in an arbitrary frame, 
a strong function of the difference 
$m \, ({\pmb V}_{\rm s}-{\pmb V}_{\rm q}) \equiv m \, \Delta {\pmb V}$.
We will refer to this effect as to the
`$\Delta {\pmb V}$-effect'.
%
\begin{figure}
    \begin{center}
        \leavevmode
        \epsfxsize=3.3in \epsfbox{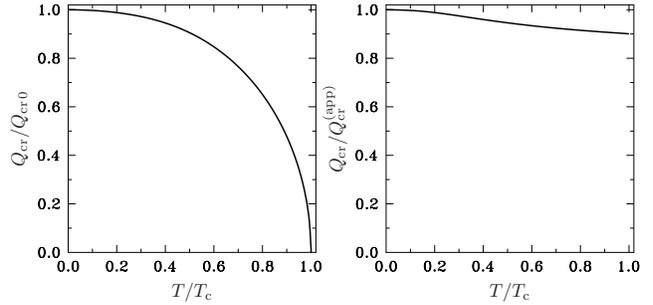}
    \end{center}
    \caption{Left panel: $Q_{\rm cr}$ (in units of $Q_{\rm cr \, 0}$)
    versus $T$ (in units of $T_{\rm c}$).
    Right panel: The same as in the left panel but 
    $Q_{\rm cr}$ is in units of $Q_{\rm cr}^{({\rm app})}$ 
    [see Eq.\ (\ref{Qcrapp})].
    }
    \label{Fig:Qcr_T}
\end{figure}
%
The critical value $\Delta V_{\rm cr}(T)$ 
of $\Delta V = |{\pmb V}_{\rm s}-{\pmb V}_{\rm q}|$,
at which superfluidity dies out, 
is easily estimated 
by taking
$Q_{\rm cr} \sim Q_{\rm cr}^{\rm (app)}$.
Then, from Eq.\ (\ref{Qcrapp}), we obtain
\begin{equation}
\Delta V_{\rm cr}(T) \sim 10^7 \left[\frac{\Delta(T, \, 0)}{10^9 \, {\rm K}}\right] \, 
\left(\frac{n_0}{n} \right)^{1/3} \,\,\, {\rm cm} \,\, {\rm s}^{-1},
\label{VsVq}
\end{equation}
where $\Delta(T,\, 0)$ is measured in Kelvins; 
$n_0=0.16$~fm$^{-3}$ is the nucleon density in atomic nuclei;
$n=p_{\rm F}^3/(3 \pi^2)$ is the particle number density.

\section{Importance of the $\Delta {\pmb V}$-effect
for neutron stars}

If the difference $\Delta V$ between 
the baryon superfluid velocities and a normal velocity
is comparable to $\Delta V_{\rm cr}$, 
then the baryon energy gaps 
can be substantially reduced. 
A few interesting consequences 
of this `dynamical reduction' of the gaps
are discussed in the next section. 
Here we illustrate possible importance of 
the $\Delta {\pmb V}$-effect 
by considering radial oscillations 
of a nonrotating superfluid NS
whose core 
is composed of neutrons, protons, and electrons.
The main question
is at what oscillation amplitude
$\Delta V$ becomes
comparable to $\Delta V_{\rm cr}$?

For simplicity we 
($i$) assume that 
neutrons pair in the spin-singlet ($^1S_0$) state 
[rather than in the triplet ($^3P_2$) state] 
and ($ii$) neglect the Landau quasiparticle interaction between quasinucleons
when calculating $\Delta_{\rm n}(T,\,  {\pmb V}_{{\rm s}\rm n} -{\pmb V}_{\rm q})$
[here and below the subscripts $\rm n$, $\rm p$, and $\rm e$  
refer to neutrons, protons, and electrons, 
respectively]\footnote{
Let us remark that 
to calculate $\Delta_{\rm n}$ and $\Delta_{\rm p}$ as functions of
$({\pmb V}_{{\rm s}\rm n} -{\pmb V}_{\rm q})$ and 
$({\pmb V}_{{\rm s}\rm p} -{\pmb V}_{\rm q})$
{\it with} allowance for interactions between quasiparticles,
one should follow the derivation of \cite{gh05}.
Namely, one should self-consistently solve
equations (31) and (32) of that reference without
making assumptions (34)--(37), 
which are valid only for small
relative velocities between the superfluid and normal components.
As a result of this calculation, 
one will find that, generally,
$\Delta_{\rm n}$ (and $\Delta_{\rm p}$) is a function
of {\it both} $({\pmb V}_{{\rm s}\rm n} -{\pmb V}_{\rm q})$ 
and 
$({\pmb V}_{{\rm s}\rm p} -{\pmb V}_{\rm q})$.
This interesting property is a consequence of entrainment between baryons
of different species; 
we will discuss it in more detail elsewhere.
}.

The NS model used here and all the microphysics input are essentially 
the same as in \cite*{kg11}; 
we refer the reader to that work for more details.
In particular, we consider 
the star of gravitational mass $M=1.4 M_{\odot}$, 
circumferential radius $R=12.2$~km, 
central density $\rho_{\rm c}=9.26 \times 10^{14}$~g~cm$^{-3}$,
and adopt the APR EOS in the NS core (\citealt*{apr98}).
The model of nucleon superfluidity employed here 
coincides with the model 3 of \cite{kg11} 
and is shown in Fig.\ 3.

The left panel of Fig.\ 3 presents nucleon critical temperatures 
$T_{{\rm c}\rm n}$ and $T_{{\rm c}\rm p}$
versus density $\rho$ in the NS core, 
the right panel demonstrates
the red-shifted critical temperatures 
$T_{{\rm c}\rm n}^{\infty}\equiv T_{{\rm c}\rm n} \, {\rm e}^{\nu/2}$
and $T_{{\rm c}\rm p}^{\infty}\equiv T_{{\rm c}\rm p} \, {\rm e}^{\nu/2}$ 
($\nu$ is the metric function)
versus radial stellar coordinate $r$ 
(in units of $R$). 
The redshifted proton critical temperature 
is taken to be constant 
$T_{\rm cp}^\infty = 5 \times 10^9$~K;
the redshifted neutron critical temperature 
varies with $r$ and has maximum 
$T_{{\rm c n} \, {\rm max}}^\infty = 6 \times 10^8$~K 
in the stellar centre.
In the right panel of Fig.\ 3 
we hatch the region occupied by the neutron superfluidity
at a redshifted stellar temperature 
$T^\infty \equiv T \, {\rm e}^{\nu/2} =4 \times 10^8$~K.
%
\begin{figure}
    \begin{center}
        \leavevmode
        \epsfxsize=3.3in \epsfbox{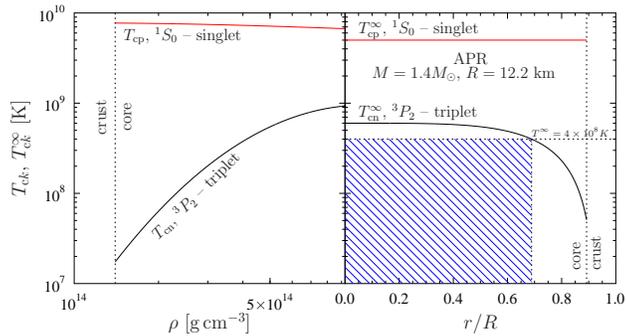}
    \end{center}
    \caption{(color online) Left panel: Nucleon critical temperatures $T_{{\rm c}k}$
    ($k={\rm n,p}$) versus density $\rho$. 
    Right panel: Redshifted critical temperatures 
    $T_{{\rm c}k}^\infty$ 
    versus radial coordinate $r$. See text for details.
    }
    \label{Fig:Tc}
\end{figure}
%

To model oscillations of superfluid NSs one has to use 
the hydrodynamics of mixtures of superfluid Fermi-liquids 
(\citealt*{ab75,ac07,ga06,gusakov07}). 
The important parameter of such hydrodynamics 
is the so called entrainment matrix $\rho_{ik}$ 
(\citealt*{ab75,bjk96,gh05}), 
or relativistic entrainment matrix $Y_{ik}$ 
(\citealt{gkh09a,gkh09b}).
Both these matrices are very temperature-dependent (\citealt{gh05,gkh09b}).
As a consequence, the eigenfrequencies and eigenfunctions
of oscillating superfluid NS 
also depend on temperature 
(\citealt{ga06,kg11}, \citealt*{cg11})%
\footnote{In this paper we use the standard (textbook) version
of superfluid hydrodynamics in which the independent velocity fields are 
${\pmb V}_{{\rm s}\rm n}$, ${\pmb V}_{{\rm s}\rm p}$, and ${\pmb V}_{\rm q}$.
Notice, however, that in the NS literature an equivalent form 
of superfluid hydrodynamics is often used which follows from the 
convective variational principle formulated by Carter 
and analyzed, in the nonrelativistic framework,
by \cite{prix04}. 
In this hydrodynamics (and in the context of npe-matter) 
the independent velocity fields
are ${\pmb v}_{l} \equiv {\pmb J}_l/\rho_l$, 
where ${\pmb J}_l$ and $\rho_l$ 
are, respectively, 
the mass current density 
and density for particle species 
$l={\rm n}$, p, and e.
These velocities are related with
${\pmb V}_{{\rm s}\rm n}$, 
${\pmb V}_{{\rm s}\rm p}$, 
and ${\pmb V}_{\rm q}$
by the following equations ($i=$n, p):
$\rho_{i}{\pmb v}_{i}= \sum_{k={\rm n}, 
\, {\rm p}} \rho_{ik} {\pmb V}_{{\rm s} k} 
+ (\rho_{i}-\sum_{k={\rm n}, 
\, {\rm p}} \rho_{ik}) {\pmb V}_{\rm q}$; 
${\pmb v}_{{\rm e}}={\pmb V}_{\rm q}$.
See \cite{prix04} for more details.
In terms of velocities
${\pmb v}_{\rm n}$, ${\pmb v}_{\rm p}$, and ${\pmb v}_{\rm e}$ 
the difference 
$\Delta {\pmb V}_{\rm n}= {\pmb V}_{\rm sn}- {\pmb V}_{\rm q}$
equals: 
$\Delta {\pmb V}_{\rm n}= 
[\rho_{\rm n} \rho_{\rm pp} ({\pmb v}_{\rm n}-{\pmb v}_{\rm e})
-\rho_{\rm p}\rho_{\rm np}({\pmb v}_{\rm p}-{\pmb v}_{\rm e})]
/(\rho_{\rm nn}\rho_{\rm pp}-\rho_{\rm np}^2)$.
}.
Below we consider the {\it first} 
radial oscillation mode of a superfluid NS
(see \citealt{kg11}, particularly figure 3 there).
%
\begin{figure}
    \begin{center}
        \leavevmode
        \epsfxsize=3.3in \epsfbox{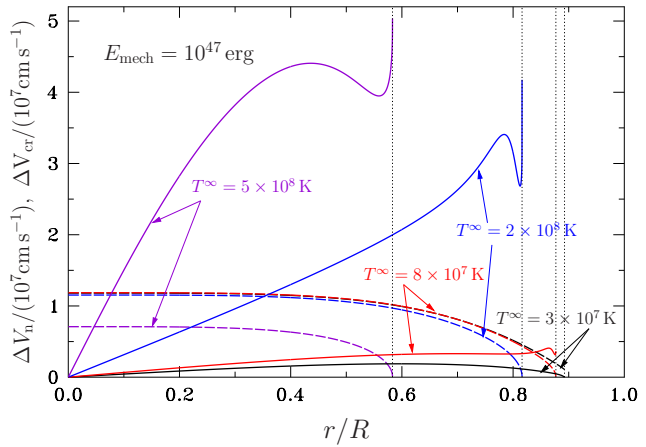}
    \end{center}
    \caption{(color online) Amplitudes of the eigenfunctions $\Delta V_{\rm n}$ 
    (solid lines) 
    and the critical velocities $\Delta V_{\rm cr}$ (dashes) 
    versus $r/R$ for the four temperatures 
    $T^\infty=3.0 \times 10^7$~K (black lines), 
    $8.0 \times 10^7$~K (red lines), $2.0 \times 10^8$~K (blue lines), 
    and $5.0 \times 10^8$~K (violet lines). 
    To plot $\Delta V_{\rm n}$ we assumed 
    that the energy of oscillations is $E_{\rm mech}=10^{47}$~erg.
    The vertical dotted lines show the (temperature-dependent) 
    boundaries between the inner superfluid and the outer normal regions.
    See text for details.
    }
    \label{Fig:dV_r}
\end{figure}

Figure 4 shows the amplitude of the eigenfunction 
$\Delta V_{\rm n} \equiv |{\pmb V}_{{\rm s}\rm n} -{\pmb V}_{\rm q}|$
and the critical velocity 
$\Delta V_{\rm cr}$ 
as functions of $r$
(solid and dashed lines, respectively; both in units of $10^7$~cm~s$^{-1}$)%
\footnote{We stress that {\it both} velocities 
${\pmb V}_{{\rm s}\rm n}$ and ${\pmb V}_{\rm q}$
are calculated self-consistently using 
the finite temperature superfluid hydrodynamics.}.
We plot $\Delta V_{\rm n}$ and $\Delta V_{\rm cr}$ 
for four redshifted stellar temperatures:
$T^\infty=3.0\times 10^7$~K (black lines), $8.0 \times 10^7$~K (red lines), 
$2.0 \times 10^8$~K (blue lines), and $5.0 \times 10^8$~K (violet lines).
The oscillation frequencies $\omega$ of the first radial mode
for such temperatures are
$\omega/(10^4 \, {\rm s^{-1}})\approx 1.702$, $1.702$, 
$1.064$, and $0.516$, respectively.

The vertical dotted lines in Fig.\ 4 indicate (temperature-dependent) 
boundaries between the neutron superfluid region 
and the outer normal region with nonsuperfluid neutrons.
In the normal region 
the functions $\Delta V_{\rm n}$ and $\Delta V_{\rm cr}$ are not defined.
The oscillation energy 
of the star is $E_{\rm mech}=10^{47}$~erg. 
For a {\it nonsuperfluid} NS
this energy corresponds to an oscillation amplitude 
\begin{equation}
\varepsilon \equiv \lim_{r \rightarrow 0} \, \frac{\xi(r)}{r}
\approx 4.4 \times 10^{-4}, 
\label{ampl}
\end{equation}
where $\xi(r)$ is the Lagrangian displacement (\citealt*{gyg05}).

It follows from Fig.\ 4 
that $\Delta V_{\rm n}$ can substantially exceed 
the critical values $\Delta V_{\rm cr}$, 
so that superfluidity is destroyed 
by oscillations in the large part of the stellar core 
(see, in particular, the violet and blue curves).
This means that the $\Delta {\pmb V}$-effect 
can greatly influence (or even drive) the dynamics of NSs 
already at rather modest oscillation amplitudes.
%
\begin{figure}
    \begin{center}
        \leavevmode
        \epsfxsize=2.5in \epsfbox{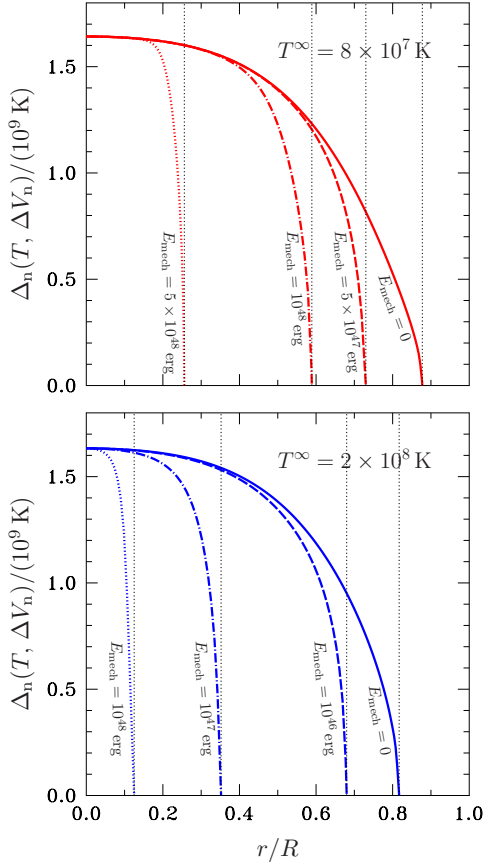}
    \end{center}
    \caption{(color online) 
    Neutron energy gap $\Delta(T, \, \Delta V_{\rm n})$ 
    (in units of $10^9$~K)
    versus $r/R$ for two temperatures $T^\infty = 8 \times 10^7$~K (upper panel)
    and $T^\infty=2 \times 10^8$~K (bottom panel) and some oscillation 
    energies $E_{\rm mech}$ (indicated in the figure). 
    Vertical dotted lines show $r$ at which neutron superfluidity disappears ($\Delta_{\rm n}=0$).
    The larger $E_{\rm mech}$ the smaller the superfluid region. 
    See text for details.
    }
    \label{Fig:Gap_r}
\end{figure}
%

This point is additionally illustrated in Fig.\ 5,
where we plot the neutron energy gap 
$\Delta_{\rm n}(T, \, \Delta V_{\rm n})$ versus $r/R$ 
for two temperatures, 
$T^\infty = 8.0 \times 10^7$~K (upper panel) and  
$T^\infty =2.0 \times 10^8$~K (bottom panel), 
and a set of oscillation energies $E_{\rm mech}$.
In the upper panel 
$\Delta_{\rm n}(T, \, \Delta V_{\rm n})$
is shown for $E_{\rm mech} = 0,\, 5.0\times 10^{47},\, 10^{48}$, and $5\times 10^{48}$~erg;
in the bottom panel 
$\Delta_{\rm n}(T, \, \Delta V_{\rm n})$
is shown for $E_{\rm mech} = 0,\, 10^{46},\, 10^{47}$, and $10^{48}$~erg.
The oscillation amplitudes $\varepsilon$
[given by Eq.\ (\ref{ampl})]
for these oscillation energies
are presented in Table.

Notice that in each panel of Fig.\ 5 the curves are plotted using 
the eigenfunctions $\Delta V_{\rm n}(r)$, 
which differ from one another only by normalization
(by the value of $E_{\rm mech}$).
For $E_{\rm mech}=10^{47}$~erg these eigenfunctions
have already been presented in Fig.\ 4
(see the red and blue solid lines; 
the red line corresponds to 
$T^\infty = 8.0 \times 10^7$~K, 
the blue line -- to $T^\infty = 2.0 \times 10^8$~K).

If $E_{\rm mech} = 0$ 
(no oscillations; see the solid lines in both panels of Fig.\ 5)
the gap $\Delta_{\rm n}$ is unaffected by $\Delta V_{\rm n}$
and is entirely determined 
by the dependence of $T_{\rm cn}$ on $r$ (see Fig.\ 3).
The vertical dotted lines in Fig.\ 5 indicate boundaries between 
the inner superfluid and the outer normal regions;
these boundaries depend on $E_{\rm mech}$. 
Obviously, the higher $E_{\rm mech}$,
the larger $\Delta V_{\rm n}(r)$, 
and, correspondingly, 
the smaller the superfluid region and $\Delta_{\rm n}$.
One sees that the gaps are very sensitive 
to variation of $\Delta V_{\rm n}$.

%
%
\begin{table}
\caption{Oscillation (mechanical) energy $E_{\rm mech}$ and the corresponding amplitude
of oscillations $\varepsilon$, defined by Eq.~(\ref{ampl}).
}
\begin{center}
\begin{tabular}{|l|c|c|c|c|c|c|c|}
\hline
$E_{\rm mech}/(10^{47} \, {\rm erg})$ & $0.0$ & $0.1$ & $0.5$ &  $1$ & $5.0$ & $10.0$ & $50.0$  \\ 
 \hline
$\varepsilon/10^{-4}$  & $0.0$ & $1.4$ & $3.1$ & $4.4$ & $9.7$ & $14$ & $31$\\ \hline
\end{tabular}
\label{tab}
\end{center}
\end{table}

\section{Possible applications}

As follows from the consideration of the previous section,
the $\Delta {\pmb V}$-effect can operate 
at not too small oscillation amplitudes.
All interesting consequences of this effect 
are related to the reduction of baryon gaps.
Let us list some of them:

(1) The reduction of the gaps
influences the entrainment matrix $\rho_{ik}$ (\citealt{gh05}), 
which depends on them.
As a result, 
$\rho_{ik}$ will become a non-linear function
of the oscillation amplitude.
This will 
($i$) make the oscillation equations nonlinear 
and hence 
($ii$) affect the eigenfrequencies and eigenfunctions of oscillating NS.
Moreover, this will 
($iii$) influence the dissipation processes, 
because bulk viscosity terms
explicitly depend on $\rho_{ik}$.
In a rotating star the decrease 
of the element $\rho_{\rm np}$ of the entrainment matrix
will, in addition, 
($iv$) reduce the mutual friction force,
which is proportional to $\rho_{\rm np}$ (\citealt*{als84}). 
We emphasize that the dependence of $\rho_{\rm np}$ on $T$ 
and on $\Delta V_{\rm n}$ and $\Delta V_{\rm p}$ 
is a very important effect for mutual friction and related phenomena, 
which has been neglected in the literature.

How to calculate the entrainment matrix $\rho_{ik}$
taking into account the $\Delta{\pmb V}$-effect?
A direct calculation is difficult 
(but one can perform it in a manner similar 
to how it was done in \citealt{gh05}).
A good approximation for $\rho_{ik}$ 
could be to calculate it from the formula (49) of \cite{gh05}
making use of the velocity-dependent gaps from Sec.\ II
instead of the gaps 
$\Delta_{\rm n}(T, \, 0)$ and $\Delta_{\rm p}(T, \, 0)$.
In this way one would obtain, for instance, for $\rho_{\rm np}$
\begin{equation}
\rho_{\rm np} = \frac{p_{\rm F n}^{3/2} p_{\rm F p}^{3/2}}{9 \,\pi^2 \, S} 
\, \frac{m_{\rm n}m_{\rm p}}{\sqrt{m_{\rm n}^\ast m_{\rm p}^\ast}}\, 
F_1^{\rm np} \,(1-\Phi_{\rm n})\,(1-\Phi_{\rm p}),
\label{rhonp}
\end{equation}
where $S = ( 1 + F_1^{\rm nn} \, \Phi_{\rm n}/3) \,
( 1 + F_1^{\rm pp} \, \Phi_{\rm p}/3 )
- ( F_1^{\rm np}/3 )^2 \, \Phi_{\rm n} \Phi_{\rm p}$;
$m_i$, $m_{i}^\ast$, $p_{{\rm F}i}$, and 
$F_1^{ik}$ 
are the mass of a free particle, 
Landau effective mass, Fermi momentum and 
the dimensionless Landau parameters, respectively 
($i,\, k={\rm n},\, {\rm p}$).
Further, $\Phi_i$ is a simple function of $x_i\equiv \Delta_i(T, \, \Delta V_{i})/T$,
specified in \cite{gh05}, 
which changes from 0 at $T=0$ to 1 at $\Delta_i(T, \Delta V_{i})=0$.
One sees from Eq.\ (\ref{rhonp}) that $\rho_{\rm np}$ vanishes
whenever $\Delta_{\rm n}(T, \, \Delta V_{\rm n})=0$ (and hence $\Phi_{\rm n}$=1)
or $\Delta_{\rm p}(T, \, \Delta V_{\rm p})=0$ (and hence $\Phi_{\rm p}=1$).

(2) Another important consequence of the $\Delta{\pmb V}$-effect
is its impact on kinetic coefficients of NS matter, 
in particular, on the bulk and shear viscosities.

($i$){\it Bulk viscosity}.
There are four bulk viscosity coefficients in the $\rm npe$-matter
of NSs (\citealt{gusakov07}). 
All of them are generated by nonequilibrium beta-processes 
(direct or modified URCA reactions)
and depend on the difference $\Delta \Gamma$ 
between the direct and inverse reaction rates.
$\Delta \Gamma$ is generally a complicated function of $T$, $\Delta_{\rm n}$, $\Delta_{\rm p}$, 
and of the imbalance of chemical potentials 
$\delta \mu \equiv \mu_{\rm n}-\mu_{\rm p}-\mu_{\rm e}$ (\citealt*{hly00,hly01}),
where $\mu_i$ is the chemical potential for particle species $i=\rm n$, $\rm p$, $\rm e$.
Recently it has been shown by \cite*{ars12}, 
that if $\delta \mu > {\rm max}\{\Delta_{\rm n}, \, \Delta_{\rm p}\}$
then, even for $T\ll \Delta_{\rm n}$ and/or $\Delta_{\rm p}$,
the bulk viscosity is {\it not} suppressed by the nucleon superfluidity 
and can be very efficient.
It seems that the $\Delta{\pmb V}$-effect of the reduction 
of the energy gaps $\Delta_{\rm n}$ and $\Delta_{\rm p}$
by relative motion of superfluid and normal component 
is {\it complementary} to the effect considered in \cite{ars12}.
Both effects act in unison to increase the bulk viscosity coefficients, 
and they are of comparable strength. 
Notice, however, that the effect of \cite{ars12} 
can only affect the bulk viscosity coefficients, 
while the applicability range of the $\Delta{\pmb V}$-effect is wider; 
it directly influences the baryon energy gaps and thus all dynamics of NSs.

($ii$) {\it Shear viscosity}.
Neglecting entrainment between baryon species ($\rho_{\rm np}=0$),
the shear viscosity $\eta$ can be calculated 
in the same fashion as was done, 
e.g., in \cite*{sy08} 
[the results will be the same].
The only difference is that one should use the velocity-dependent 
gaps $\Delta_i(T, \, \Delta V_i)$ 
instead of $\Delta_i(T,\, 0)$ in all equations 
[$i=\rm n$, $\rm p$].
It is interesting that the $\Delta {\pmb V}$-effect can both 
increase or decrease the shear viscosity. 
For example, the electron shear viscosity $\eta_{\rm e}$ decreases 
with increasing $\Delta V_{\rm p}$ 
(that is, with reducing $\Delta_{\rm p}$), 
because electrons are better screened by protons 
when $\Delta_{\rm p}$ is large (\citealt{sy08}).
On the other hand, 
the neutron shear viscosity $\eta_{\rm n}$ 
can either decrease or increase 
with growing $\Delta V_{\rm n}$ and $\Delta V_{\rm p}$.
The behaviour of $\eta_{\rm n}$
in that case is determined by the competition of two effects:
by the increase of the normal density 
of neutron Bogoliubov excitations $\rho_{\rm qn}$
and by the reduction of the neutron mean free path $\lambda$ 
due to more frequent collisions with neutron and proton Bogoliubov excitations
(note that $\eta_{\rm n}$ can be estimated as 
$\eta_{\rm n} \sim \rho_{\rm q n} \, v_{\rm Fn} \, \lambda$, 
where $v_{\rm Fn}$ is the neutron Fermi-velocity).
Similar effects were carefully analyzed in \cite*{bhy01} 
in application to the neutron thermal conductivity.

An entrainment between neutrons and protons
will strongly modify the derivation of the 
neutron shear viscosity, 
even neglecting the $\Delta {\pmb V}$-effect.
The main difference will be the equilibrium 
Fermi-Dirac distribution function 
for neutron Bogoliubov excitations in a system
with superfluid currents.
This function was first obtained in \cite{gh05} 
[see equation (28) there]; 
it is very different from the standard expression, 
valid when $\rho_{\rm np}=0$.
To our best knowledge, 
a derivation of $\eta_{\rm n}$ in a system {\it with} entrainment 
has not been attempted in the literature.

(3) Finally, there is a number of important consequences of the fact
that the relative velocity $\Delta {\pmb V}$ 
between the superfluid and normal liquid components
cannot be too large in a {\it stationary} rotating NS. 
Here we present two of them.

($i$) It is generally accepted that neutron vortices are pinned 
to atomic nuclei in the NS crust 
(or to magnetic flux tubes in the NS core).
At a certain critical $\Delta {\pmb V}$ 
they can unpin from the nuclei 
(or from magnetic flux tubes).
However, 
in some models (e.g., \citealt*{link09})
pinning is so strong that 
the critical relative velocity can be 
as high as $10^6 \div 10^7$~cm~s$^{-1}$.
These values are close to 
$\Delta V_{\rm cr}$ [see Eq.\ (\ref{VsVq})].
Thus, 
the $\Delta {\pmb V}$-effect 
can be very important for such models.
It can also play a role in explanation 
of the long-period precession of isolated pulsars (\citealt{link03})%
\footnote{We thank the anonymous referee 
for pointing out to us this possibility.}.

($ii$) In \cite*{acp04} and \cite{slac10}
a two-stream instability 
in homogeneous superfluid matter
is discussed,
that can be triggered once the relative velocity
$\Delta {\pmb V}$ reaches some critical value.
According to these authors, 
the critical value is of the order of the sound speeds,
i.e., it is {\it much greater} 
than the typical $\Delta V_{\rm cr}$,
at which superfluidity completely disappears 
[see Eq.\ (\ref{VsVq})].
In other words, 
it is not very probable 
that this instability is realized in NSs.
Notice, however, 
that under certain circumstances 
similar instability in rotating NSs 
can drive the so called inertial modes
unstable at a much lower $\Delta {\pmb V}$ 
(\citealt*{pca04}).

\section{Conclusion}

The baryon energy gaps depend 
on the relative velocity between the superfluid
and normal components ($\Delta {\pmb V}$-effect). 
We propose, for the first time, 
that this effect may have a strong impact 
on the dynamical properties of NSs. 
We illustrate this point by considering 
radial oscillations of an NS 
with superfluid nucleon core
and a nonsuperfluid crust.
However, we stress that the $\Delta {\pmb V}$-effect 
should be equally important in the crust of NSs 
where superfluid neutrons are present, 
as well as in the interiors of hyperon and quark stars.
Although we discussed some immediate applications in Sec.\ IV,
it is clear that
more efforts are needed to analyze all possible
consequences of this effect
on the evolution of NSs.

\section*{Acknowledgments}

The authors are very grateful to D.G. Yakovlev
for valuable comments and encouragement.
This research was supported by  
Ministry of Education and Science of Russian Federation 
(Contract No. 11.G34.31.0001 
with SPbSPU and leading scientist G.G. Pavlov,
and Agreement No. 8409, 2012), 
RFBR (grants 11-02-00253-a and 12-02-31270-mol-a), 
FASI (grant NSh-4035.2012.2), 
and by RF president programme (grant MK-857.2012.2).

\bibliography{author}

\label{lastpage}

\end{document}